\documentclass{svjour8}                    
\smartqed  
\usepackage{graphicx,amsmath,amssymb,amsbsy,latexsym,verbatim}
\newcommand{\be}{\nopagebreak[3]\begin{equation}}
\newcommand{\ee}{\end{equation}}
\newcommand{\ba}{\nopagebreak[3]\begin{eqnarray}}
\newcommand{\ea}{\end{eqnarray}}

\newcommand{\bc}{}

\begin{document}

\title{Infinities as a measure of our ignorance}

\subtitle{Notes for the workshop ``Infinities and Cosmology"}
\author{Francesca Vidotto}
\institute{
Radboud University Nijmegen, Institute for Mathematics, Astrophysics and Particle Physics, 
Mailbox 79, P.O. Box 9010, 6500 GL Nijmegen, The Netherlands\\\email{fvidotto@science.ru.nl}
}
\thanks{The work of FV at Radboud University is supported
by a Rubicon fellowship from the Netherlands Organisation for Scientific Research (NWO).}

 \date{DAMPT, Cambridge, UK, 21st March 2013}
\maketitle

\begin{abstract}
The quantization of gravity offers a solution to the presence of singularities in cosmology. Infinities are removed because of the existence of finite quanta of spacetime. This is one of the most important prediction of Loop Quantum Gravity.
But treating gravity on the same footing as the other quantum field theories introduces a different kind of infinities: the ones from the renormalization procedure. 
In the covariant formulation of Loop Quantum Gravity (Spinfoam) this kind of infinities are tamed by the presence of the cosmological constant.
These two results are guided by a specific perspective on the appearance of infinity  in physics: this is just interpreted as a signpost that the theory should be improved. It arises from our limited knowledge, and not as a fundamental fact of nature.
 
\end{abstract}
\vskip3em
\begin{center}
\emph{To see a world in a grain of sand\\
And heaven in a wild flower,\\
Hold infinity in the palm of your\\
Hand and eternity in one hour.}\\
William Blake
\end{center}

\section{Becoming and discreteness}
The first question that we can trace in the history of Western philosophy seems to be, at the eyes of a contemporary physicist, the mother of all questions: what is Being? what is Becoming?  Are these two states conciliable?  This history, whose main  characters can be identified in Parmenides and Democritus, passes trough the figure of Zeno. Zeno was at the same time a disciple of Parmenides, but also a friend of Leucippus, the grandfather of atomism, later fully developed in Democritus. Zeno proposed his famous paradox as a tale of Achilles and the turtle. As traditionally phrased, Zeno was confused by the assumption that every infinite series diverges, while the convergence of infinite series has been studied by mathematicians only in the XVIII century. Despite the way out from the paradox provided by later mathematicians, the core of the problem addressed by Zeno still stands as an open problem in our contemporary understanding of the world. The usual interpretation of the paradox is that Zeno was trying to demonstrate the thesis of Parmenides, according to whom the fundamental unity of everything would not allow change to exist: change would be an illusion. But anther possible interpretation could instead place Zeno on the opposite side, the side of his friend Leucippus: we see change, and in order not to fall in the paradoxical paralysis of Achilles, we have to assume that the word is made by fundamental discrete pieces, the Democritean atoms.  
In this view, what in fact exists is only the becoming, that in the language of modern physics we would call \emph{process}. 
 And this is intimately entangled to the fundamental discrete nature of everything, in particular of space and time. 


Quantum gravity, at least as presented by Rovelli's school, appears as the contemporary heir of this position. 
On the one side, there is the relational interpretation of quantum mechanics, where all the ontological meaning is shifted away from the wave function in favor of the measurement outcomes,  and the disappearing of an absolute time in favor of many relational times governed by the interactions, as suggested by general relativity.
On the other side, we have the basic result of Loop Quantum Gravity \cite{Rovelli:1989za,Rovelli,Introduction} : the theory \emph{predicts} the existence of quanta of spacetime. In Zeno's words, we cannot subdivide space infinitely. 
There exist a fundamental scale, the Planck length.

\section{Constants of Nature}

The existence of a minimal length scale is the main feature of quantum gravity and gives it a universal character, analogous to special relativity and quantum mechanics.  Special relativity can be seen as the discovery of the existence of a maximal local physical velocity, the speed of light $c$.  Quantum mechanics can be interpreted as the discovery of a minimal action $\hbar$ in all physical interactions, or equivalently the fact that a finite region of phase space contains only a finite number of distinguishable (orthogonal) quantum states, and therefore there is a minimal amount of information in the state of a system. Quantum gravity is the discovery that there is a minimal length.  

The existence of a fundamental length, the absolute building block of spacetime structure, has been realized since when both Quantum Mechanics and General Relativity became available. 
This can be easily understood using some few semiclassical considerations.
Every particle is a localized excitation of some field, but the localization cannot be more precise than the intrinsic uncertainty derived from Heisenberg relation: $\Delta x >\hbar/\Delta p$, so that for a sharp location a large momentum $p$ is required.
Large momentum implies large energy $E$. In the relativistic limit, where rest mass is negligible, $E\sim c p$. Sharp localization requires large energy.  
But General Relativity teaches us that any form of energy $E$ acts as a gravitational mass $M\sim E/c^2$ and distorts spacetime around itself. The distortion increases when energy is concentrated, to the point that a black hole forms when a mass $M$ is concentrated in a sphere of radius $R \sim G M/c^2$, where $G$ is the Newton constant. As a consequence, if we concentrate a lot of energy in $\Delta x$ to get a sharper localization, the horizon radius $R$ will grow to the point to be larger then $\Delta x$. But in this case the region of size $\Delta x$ that we wanted to mark will be hidden beyond a black hole horizon, preventing the localization.  Therefore it has no mean to decrease $\Delta x$ lower than $\Delta x=R$. 
Combining the relations above, one realizes that there exists a minimal size where a quantum particle cannot be localized without being hidden by its own horizon.
This is the Planck length
$$
\ell_{P}\ =\ \sqrt{\frac{\hbar G}{c^3}} \ \sim \ 10^{-35} \,m.
$$  

This is the simplest combination of Quantum Mechanics and General Relativity.
It tells us that below the Planck scale it make no sense to talk about distances. It points to the presence of a quantum discreteness but still  it does not tell us how the notion of quantum discreteness may emerge in spacetime.
People realized immediately the importance of this assumption for the resolution of cosmological singularities, but what was lacking was a mechanism to see this fundamental length emerging in a quantum theory of gravity.

The presence of a minimal length in the infinitely small has consequence for the infinitely large \cite{Bianchi:2011ys}. 
Suppose, as Einstein liked, that our universe is a 3-sphere with radius, say, $1/{\sqrt{\Lambda}}$, or that we live in a Lorentzian space with a horizon at that distance. Since we cannot observe anything smaller than the Planck length, if there is an object of size $\ell_{P}$ at such a distance, we will never see it under an angle smaller than $
\phi_{min}=\sqrt{\Lambda}\, \ell_{P}.$

In such a situation, everything we see is captured, on the local celestial 2-sphere formed by the directions around us, by spherical harmonics with $j \le j_{max}$. Since the  $j$-th spherical harmonic distinguishes dihedral angles of size $\phi^2\sim4\pi/(2j+1)$, we won't see harmonics with 
$
j>j_{max}=4\pi/\phi_{min}^2\sim {4\pi}/{\ell^2_P\, \Lambda}.
$
The quantization of geometry suggests that also angles should be quantized \cite{Major:1999mc,Major:2011ry}, pointing to the presence of another fundamental constant: the cosmological constant, an uncertainty factor in the minimal resolution of small angles in the sky.


\section{Quanta of spacetime}

The lesson of General Relativity yields background independence: spacetime is not the playground for field interactions, but is an interacting field as the others. Our modern understanding of fundamental forces, or better said fields, is that they are associated to some gauge symmetry. In the Standard Model, these symmetries are respectively $U(1)$ for electromagnetism, $SU(2)$ for the weak interaction and $SU(3)$ for the strong interaction. Gravity does not differ and can be written as a gauge theory.

General Relativity admits a version where, instead of the metric, the fundamental object  are reference fields: the tetrads. One can always express the metric in terms of the tetrads, therefore the tetrad version can be consider equivalent to the original metric formulation. On the other hand, only in the tetrad formulation it is possible to couple fermions to the gravitational field: since we observe fermions in nature, this makes the tetrad version of General Relativity somehow more fundamental.

So let us start by the tetrads. 
The invariance under diffeomorphisms of General Relativity imposes the independence by coordinate transformations. This implies that for each point of spacetime we have to associate a tetrad that is locally Lorentz invariant. But time is pure gauge in General Relativity and we can always fix this gauge, leaving us only with the rotational part of the Lorentz transformation.%
\footnote{The canonical analysis of General Relativity shows that when we perform an ADM decomposition of spacetime, the Lapse and the Shift functions are in fact Lagrangian multipliers. In the covariant theory, this is reflected in the simplicity constraint that connect respectively boost and rotation as $\vec K+\gamma \vec L=0$, where $\gamma$ is the Immirzi parameter.} 
This gauge invariance, that naturally arises in the classical gravitational theory, is the starting point in order to jump to the quantum theory.
The quantum states have to be thought as boundary states \cite{Oeckl:2003vu,Oeckl:2005bv}, describing the space geometry at some fixed time.
When it comes to quantization, the tetrad turns out to be the generator of SU(2) transformations, satisfy the well-known algebra of the angular momentum. This implies that spacetime is quantized with a discrete spectrum. So actually we don't have any more a tetrad for each spacetime point, but a tetrad for each quanta of spacetime. On each quanta of spacetime it does not matter how the reference fields are oriented, but only the relations between adjacent quanta. A spinnetwork state in Loop Quantum Gravity \cite{Rovelli:1995ac,Introduction} is a gauge invariant state (invariant under the rotations of the triads) that knows about the excitations of each quanta of spacetime (its spin, that is related to its physical size) and the adjacency relation between them (coded in an abstract graph).

The gauge invariance of the triads yields the presence of a gauge field, as in the other Yang-Mills theories. This is an object in the Lorentz algebra, that code the information about (intrinsic and extrinsic) curvature. 
In order to define gauge invariant observables in the quantum theory, we consider the path-ordered exponential of the gauge field. This is called a Wilson loop\footnote{These are Wilson loops in the language of particle physics or holonomies in the language of differential geometry: if the gauge field is seen as the connection over the $SU(2)$ principal bundle, the holonomy is its parallel transport.} (from this the designation Loop for the quantum theory) and turns out to be the canonically conjugate variable to the triad. 

Loop Quantum Gravity variables are group variables, as the variables of the other interactions are. Since $SU(2)$ is a compact group, the spectrum of the observables corresponding to these group variables are discrete. In particular, in Loop Quantum Gravity the geometry can be describes trough observables such as areas, volumes and angles, constructed starting from the operator corresponding to the triads.

The presence of a minimal eigenvalue in the discrete spectrum of the area plays in cosmology the same role of the minimal eigenvalue of the angular momentum for atomic physics. In the classical theory, all the trajectories fall into a singularity: the electrons spiral down into the atomic nucleus, all the matter of the universe gets evolved back in time into the big bang. The dissolution of continuous spacetime into quanta of spacetime prevents singularity theorems to apply.%
\footnote{The left hand side of the classical EinsteinÕs equations is modiÞed by the quantum geometry corrections, invalidating the hypothesis that Penrose and Hawking had \cite{Hawking:1969sw}. On the other hand, the more recent singularity theorems by Borde, Guth and Vilenkin \cite{Borde:2001nh}  do not refer to EinsteinÕs equations. They assume that the expansion is positive along any past geodesic as they are motivated by the eternal inßationary scenario. Because of the pre-big-bang contracting phase, this assumption is violated in the LQC effective theory.}
Singularity theorems do not point to the presence of some infinity in Nature, but they rather signal the boundary beyond which the classical theory ceases to be valid.

The Loop quantization removes all the cosmological (physical) singularities \cite{Bojowald:2001xe,Ashtekar:2009kx,Livine:2012kl}. It is a genuine consequence of the quantization, rather than the result of some exotic condition imposed to our universe. There is no fine-tuning of initial conditions, nor an ad hoc boundary condition at the singularity. Furthermore, matter can satisfy all the standard energy conditions.%
\footnote{There is a zoo of possible cosmological singularities, beyond the notorious big bang: big rip singularities, sudden singularities, big freeze singularities, big brake... When the singularity regards spacetime itself, taking the form of a divergence in the curvature or of its derivative, loop quantization promptly resolve it. If instead the singularity is a divergence in the pressure or its derivative, loop quantization seems to have nothing to say: these are not singularity where spacetime breaks as geodetics can be continued trough them \cite{Singh:2009mz,Sami:2006wj,Singh:2003au,Singh:2010qa}.} 
The result has proven to hold even in presence of anisotropies and inhomogeneities.

\section{Covariant transition amplitudes}
Quantum discreteness of spacetime is a powerful achievement, but it is only the starting point of a beautiful journey into a new quantum theory. The kinematical space of Loop Quantum Gravity has some powerful elements of novelty, with deep consequence for our understanding of space and time through cosmological abyss.
It opens the door to a new world, where new infinities must to be faced in order for the theory to survive.

The early attempts to quantize gravity with perturbative techniques got stuck%
\footnote{%
The non-renormalizability obtained in perturbation theory has pushed the theoreticians into a quest for a larger renormalizable or finite theory. The quest has wandered through the investigation of modifications of GR with curvature square terms in the action, Kaluza-Klein-like theories, supergravity, and has lead to String Theory, a presumably finite quantum theory of all interactions including gravity, defined in 10 dimensions, including supersymmetry and so far difficult to reconcile with the observed world.
}
because of the non-renormalizability of the resulting theory. In a non-perturbative approach to quantum gravity, this problem can be overcome having clear in mind that the existence of the Planck length sets quantum gravity aside from standard quantum field theory for two reasons.  First, we cannot expect quantum gravity to be described by a local quantum field theory, in the strict sense of this term \cite{Haag:1992hx}. Local quantum theory requires quantum fields to be described by observables at arbitrarily small regions in a continuous manifold. This is not going to happen in quantum gravity. 
Second, the quantum field theories of the standard model are defined in terms of an \emph{infinite} renormalization group. The existence of the Planck length indicates that this is not going to be the case for quantum gravity.

When computing transition amplitudes for a field theory using perturbation methods, infinite quantities  appear. What is the nature of these infinities? 
Perturbation methods are some kind of approximation. 
Infinities arise because we perturb around points that are not really good.%
\footnote{If we fly from Paris to Toronto and at the first order we disregard the resistance of the air while studying the trajectory, we would find that at the first order the airplane is on Saturn, so we have to perform the normalization and bring it back.}
In Feynman graphs the interaction for small loops take us to an arbitrarily small dimension, but this is not a physical fact: it is a mistake of the approximation used.
Renormalization is a powerful tool for the computation, but it does not mean that every time that there is a phenomenon at some scale there are all the infinities down of that scale. In fact, renormalization theory is exactly what tells us that we can correct all that just by readjusting the coefficients. Even in a renormalizable Quantum Field Theory, where we need to compute only a finite number of (properly chosen) degrees of freedom, renormalization has to be used.

The standard technique consists in the introduction of a cut-off which removes the infinities. The definition of the theory is adjusted so that the cut-off dependence tune the physical observables to match the experimental observations.  The cut-off can be regarded as a technical trick, not something physical. Accordingly, care is taken so that the final amplitudes do not depend on the short scale cut-off.%
\footnote{Condensed matter offers a prototypical example of independence from short-scale cut off: second order phase transitions.  At a critical point of a second order phase transition, the behavior of the system becomes scale independent, and large scale physics is largely independent from the microscopic dynamics.}
This general structure has proven effective for describing particle physics, but it is not likely to be the structure that  works for quantum gravity.   

In quantum gravity the cut off in the modes is not a mathematical trick for removing infinities, but a genuine physical feature of the quantum spacetime.%
\footnote{This is not a strange situation: on the contrary, it is usual condensed matter away from the critical point. Take, for instance, a bar of iron at room temperature.  Its behavior at macroscopic scales is described by a low energy theory, characterized by a certain number of physical constants.  This behavior includes wave propagation and finite correlation functions. The tower of modes of the bar has an effect on the value of the macroscopic physics, and can still be explored by studying the renormalization group equation describing the dependence of physical parameters on the scale.  But the system is characterized by a physical and \emph{finite} cut-off scale, the atomic scale; and there are no modes of the bar beyond this scale.  The bar can be described as a system with a large but \emph{finite} number of degrees of freedom.} 
The Planck length provides an ultraviolet physical cut off. The cosmological constant provide an infrared cut off. These make the theory finite. 

\vskip1em

How does the cosmological constant enter the definition of the theory? The presence of a cosmological constant, as Einstein's theory allows and as experiments confirm, implies the presence of an horizon, a maximal distance. We have seen that this, combined with the existence of a minimal length, implies that we can not resolve small angles on the local celestial 2-sphere formed by the directions around us. This  physical situation is realized by the notion of ``fuzzy sphere" \cite{Madore:1991bw,Madore:1997ta,Madore:2002fk,Freidel:2001kb}, described by the algebra of the angular functions spanned by the spherical harmonics with   $j\le j_{max}$.
The mathematics of this kind of systems is realized by the notion of quantum group: the representations of the quantum deformation of the rotation group
 $SU_q(2)$, where $q=e^{i2\pi/k}$,  are characterized by a maximum angular momentum $k\sim 2j_{max}$  \cite{Majid:1988we,Maggiore:1993zu,Majid:2000fk}.

This picture is realized concretely in loop gravity for quantizations of general relativity with a cosmological constant (see for instance \cite{Major:1995yz,Smolin:2002sz,Freidel:1998pt}).   In Loop Quantum Gravity the angle $\phi$ between two directions in space is an operator with a discrete spectrum  \cite{Penrose2,Major:1999mc}. Its eigenvalues are labeled by two spins $j_1$, $j_2$, associated to the two directions, and a quantum number $k=|j_1-j_2|,\ldots, j_1+j_2$. The expression of the angle eigenvalues present a  minimal one, 
that yields an angular resolution no better than
$
\phi_{min}=\sqrt{2/j_{max}}\;
$ for some $j_{max}\gg 1$. 
 
The $SU_q(2)$ quantum deformation parameter $q$ is related to the Planck length $\ell_P$ and the cosmological constant $\Lambda$: $
                 q=e^{i\Lambda l^2_P}. $
The combination of the quantum-gravitational space granularity with the maximal size determined by the cosmological constant yields immediately a local quantum-group structure.
Notice that the deformation parameter is dimensionless, and by itself does not determine a scale at which physical space becomes fuzzy.

The quantum deformation affects the local gauge.  The local gauge  of General Relativity in the time gauge is $SU(2)$, interpreted as the  universal covering of the group of physical rotations around any given point in space. 
In a  universe characterized by an horizon at distance $1/{\sqrt{\Lambda}}$ and a minimal length $l_P$, the local rotational symmetry is better described by $SU_q(2)$ than by $SU(2)$. A quantum deformation of this  group corresponds, physically, to a non-commutativity, and therefore to a consequent intrinsic fuzziness, of any angular function. In other words, it describes the impossibility of resolving small dihedral angles of view. 

The use of quantum groups in Loop Quantum Gravity started with \cite{E.Buffenoir:kx,Noui:2002ag} and has gained a central role in the covariant approach \cite{Han:2010pz,Fairbairn:2010cp,Han:2011nx}.
On the other hand,
quantum groups and non-commutative spaces have been repeatedly utilized in various approaches to quantum gravity \cite{Snyder:1946qz,Chamseddine:1992yx,
Jevicki:2000it,Moffat:2000gr,Vacaru:2000yk,Cacciatori:2002ib,Cardella:2002pb,Vassilevich:2004ym,Buric:2006di,Abe:2002in,Valtancoli:2003ve,Kurkcuoglu:2006iw,Krajewski:1999bg,Grosse:2004yu,Freidel:2005me,Szabo:2006wx,
Livine:2008hz}. The associated non-commutative spaces are generally interpreted as describing a Planck-scale quantum uncertainty in position, while from this prospective the fuzziness is instead in the directions,  as noticed in \cite{Bianchi:2011ys}. This geometrical interpretation is compatible with the spectral point of view on space implicit in \cite{albook} and \cite{AmelinoCamelia:2011bm} and with the 
4d-angle (speed) quantization in \cite{Girelli:2003az}.


\subsection{A finite quantum cosmology}
	A system manifests its quantum nature in different ways. Quantum mechanics is often popularized as a theory about the world at short distance, but a more careful understanding would make us realize that $\hbar$ may enter in the description of large system, till the extreme case of the description of universe as a whole.
	
The quantization may appear as the discretization of physical quantities, associated to a short-scale ÒfuzzinessÓ implied by the uncertainty relations. 
We have seen that the quantization of spacetime implies the discretization into quanta of spacetime. As the mathematical tools of continuos differential geometry are substituted with discrete analogues, the insidious infinities such as the cosmological singularities disappear.

Another manifestation of the quantization results in the probabilistic nature of the evolution of the system, so that the evolution has to be expressed by  transition amplitudes. We have seen that Loop Gravity provides \emph{finite} transition amplitudes between spacetime states.
In particular, one can calculate the probability to go from an universe with a given geometry to another \cite{Bianchi:2010zs,Bianchi:2011ym,Vidotto:2011qa}. The states can describe a homogeneous and isotropic geometry, or states with inhomogeneities and anisotropies \cite{Rennert:2013pf,Alesci:2013xd}. 

The removal of singularities and the availability of states for the geometry of the universe, that are genuinely quantum-gravity states, open the investigation of a new physical region. Where before there were infinities and physical non-senses, now there are physical states that can be studied. 
Finally, philosophical issues such as the \emph{initial conditions} or the \emph{cosmological averaging} find a new promising framework where they can be addressed.

\section{Holy infinities}
\label{intro}
According to Eliade's definition of sacred \cite{Eliade1957}, sacred is everything that we feel such as not belonging to human world. It could be a space, it could be a time, it could be everything that goes beyond our human experience. In this world, we size things with respect to us: we say heavens are up in the sky because we are confined down by gravity, we measure land in terms of our foot, we count as much as needed by our daily experience. Modeling, sizing and counting are sophisticated tools that we have developed but are not given a priori. There are still some of our brothers in Amazonia and in Australia, in whose culture does not exist a word for \emph{four}: they know \emph{one}, they know \emph{two}, they know \emph{three}, then they have a word for \emph{many}, and one for \emph{many many}. Some of us can count longer, but this pose us a problem: is there an end to counting? Is there something that we can not model/size/count?
We call infinite what is beyond our human experience. We call in this way what cannot be said. But if this is a condition of our human nature, it has not to be something that has necessarily to exist in Nature. 
Infinity can exist as a mathematical object, a powerful tool for our calculation. But physics is not just mathematics. It is a discourse about Nature trough the mathematical language, where the ultimate goal is to associate a number to a physical system, and from this number a meaning. An infinity has the meaning of \emph{beyond our present knowledge}. But making science and making physics is to constantly push forward the boundary of our knowledge, in a process that cannot have an end. 

When an infinity is presented to a scientist, the scientist's job is to find a theory that replace such an infinity with a finite number.
\emph{``There are some, king Gelon, who think that the number of the sand is infinite in multitude"} starts Archimedes' ``Sand Reckoner" \cite{Sand1956}.
Archimedes engages in the count  of how many grain of sands there would be if the whole universe would have been filled by them. In order to picture what no other mind had been able to picture before, he invents a new notations for big numbers, in a way not so distant from the current scientific notation. He was able to name what before had no other name but a generic ``infinite". A prototypical scientific deed.

But sacred text still beware us about infinities: they belong to the sacred, they should not be addressed by humans.
\begin{center}
	\emph{
	\noindent 
	All wisdom comes from Yahweh \\
	and with him it remains forever.\\
	The sand of the seashore, and the drops of rain, \\
	and the days of eternity: who can number these?\\
	Heaven's height, earth's breadth, \\
	the depths of the abyss: who can explore these?\\
	Before all other things wisdom was created;\\
	and prudent understanding, from eternity.\\
	The wellspring of wisdom is the word of God in the heights,\\
	and her runlets are the eternal commandments.}\\
	Joshua ben Sira \cite{Sirach}
\end{center}
Despite centuries of rational investigation, are we still under the spell of the sacrality of infinities?

The history of modern cosmology is an example of this tension between the rational explanation of the world and find the holy place where gods hide. 
We due the understanding that Einstein's equation were bearing an unstable universe, coming from an expanding history, to the Belgian priest George Lema\^itre. He was the one giving meaning to Friedmann's calculation, telling Einstein that such an instability was unavoidable (with or without the cosmological constant), and finding experimental support of his theory in galaxy redshift.
A beginning for the universe, was seen by part of the clergy as opportunity to use of science to support religious beliefs (Genesis' \emph{fiat lux}), at the point that Pope Pius XII publicly speck about the big bang as a good manifestation. 
Lema\^itre reacted to this: he did not believe that any scientific true should be searched in the Bible, as scientific opinions in the Bible reflect only the common knowledge at the time of writing. He was aware of the precarious condition of every scientific truth: today's infinities may become finite tomorrow... 

\section{Conclusions}
Physics is about a quantitative description of the world around us. Infinities represent an end point for physics: their appearance means that we can  not associate a finite number to the system under study. Infinities provide the most tantalizing paradoxes to physicists: they point to the old assumptions that we have to give away in order to go beyond our present theories, toward the explorations of new level of energy, space and time. In this sense, every infinity and its overcoming is the turning point for a new paradigm.
\vskip2em

\centering
``I think that what is truly infinite may just be the abyss of our ignorance."
Carlo Rovelli \cite{Infinity2011}
\footnote{That is a fair way to rephrase the famous sentence by Einstein
``Two things are infinite: the universe and human stupidity; and I'm not sure about the universe."}
\vskip3em

\begin{acknowledgements}
I certify that what exposed above is fully Rovellian.\\
The work of FV at Radboud University is supported
by a Rubicon fellowship from the Netherlands Organisation for Scientific Research (NWO).
\end{acknowledgements}

\bibliographystyle{hunsrt} 
\bibliography{bib,phil,BiblioCarlo}

\begin{thebibliography}{10}

\bibitem{Rovelli:1989za}
Carlo Rovelli and Lee Smolin.
\newblock {Loop Space Representation of Quantum General Relativity}.
\newblock {\em Nucl. Phys.}, B331:80, 1990.

\bibitem{Rovelli}
Carlo Rovelli.
\newblock {\em Quantum Gravity}.
\newblock Univ. Pr., Cambridge, UK, 2004.

\bibitem{Introduction}
Francesca~Vidotto Carlo~Rovelli.
\newblock {\em Explorations in Quantum Gravity. An introduction to Covariant
  Loop Gravity}.
\newblock 2013.

\bibitem{Bianchi:2011ys}
Eugenio Bianchi and Carlo Rovelli.
\newblock A note on the geometrical interpretation of quantum groups and
  non-commutative spaces in gravity.
\newblock {\em Phys.Rev.D}, 84:027502, 2011, 1105.1898.

\bibitem{Major:1999mc}
Seth~A. Major.
\newblock {Operators for quantized directions}.
\newblock {\em Class.Quant.Grav.}, 16:3859--3877, 1999, gr-qc/9905019.

\bibitem{Major:2011ry}
Seth~A. Major.
\newblock {Quantum Geometry Phenomenology: Angle and Semiclassical States}.
\newblock {\em J.Phys.Conf.Ser.}, 360:012061, 2012, 1112.4366.

\bibitem{Oeckl:2003vu}
Robert Oeckl.
\newblock {A 'general boundary' formulation for quantum mechanics and quantum
  gravity}.
\newblock {\em Phys. Lett.}, B575:318--324, 2003, hep-th/0306025.

\bibitem{Oeckl:2005bv}
Robert Oeckl.
\newblock {General boundary quantum field theory: Foundations and probability
  interpretation}.
\newblock {\em Adv. Theor. Math. Phys.}, 12:319--352, 2008, hep-th/0509122.

\bibitem{Rovelli:1995ac}
Carlo Rovelli and Lee Smolin.
\newblock {Spin networks and quantum gravity}.
\newblock {\em Phys. Rev.}, D52:5743--5759, 1995, gr-qc/9505006.

\bibitem{Hawking:1969sw}
S.W. Hawking and R.~Penrose.
\newblock {The Singularities of gravitational collapse and cosmology}.
\newblock {\em Proc.Roy.Soc.Lond.}, A314:529--548, 1970.

\bibitem{Borde:2001nh}
Arvind Borde, Alan~H. Guth, and Alexander Vilenkin.
\newblock {Inflationary space-times are incompletein past directions}.
\newblock {\em Phys. Rev. Lett.}, 90:151301, 2003, gr-qc/0110012.

\bibitem{Bojowald:2001xe}
Martin Bojowald.
\newblock {Absence of singularity in loop quantum cosmology}.
\newblock {\em Phys. Rev. Lett.}, 86:5227--5230, 2001, gr-qc/0102069.

\bibitem{Ashtekar:2009kx}
Abhay Ashtekar.
\newblock Singularity resolution in loop quantum cosmology: A brief overview.
\newblock {\em J.Phys.Conf.Ser.}, 189:012003, 2009, 0812.4703.

\bibitem{Livine:2012kl}
Etera~R. Livine and Mercedes Mart{\'\i}n-Benito.
\newblock Group theoretical quantization of isotropic loop cosmology.
\newblock {\em Phys. Rev.}, D85:124052, 2012, 1204.0539.

\bibitem{Singh:2009mz}
Parampreet Singh.
\newblock {Are loop quantum cosmos never singular?}
\newblock {\em Class. Quant. Grav.}, 26:125005, 2009, 0901.2750.

\bibitem{Sami:2006wj}
M.~Sami, Parampreet Singh, and Shinji Tsujikawa.
\newblock {Avoidance of future singularities in loop quantum cosmology}.
\newblock {\em Phys. Rev.}, D74:043514, 2006, gr-qc/0605113.

\bibitem{Singh:2003au}
Parampreet Singh and Alexey Toporensky.
\newblock {Big crunch avoidance in k = 1 loop quantum cosmology}.
\newblock {\em Phys. Rev.}, D69:104008, 2004, gr-qc/0312110.

\bibitem{Singh:2010qa}
Parampreet Singh and Francesca Vidotto.
\newblock {Exotic singularities and spatially curved Loop Quantum Cosmology}.
\newblock {\em Phys. Rev.}, D83:064027, 2011, 1012.1307.

\bibitem{Haag:1992hx}
Rudolf Haag.
\newblock {\em Local Quantum Physics: Fields, Particles, Algebras}.
\newblock Springer, August 1996.

\bibitem{Madore:1991bw}
J.~Madore.
\newblock {The Fuzzy sphere}.
\newblock {\em Class.Quant.Grav.}, 9:69--88, 1992.

\bibitem{Madore:1997ta}
J.~Madore.
\newblock {Gravity on fuzzy space-time}.
\newblock 1997, gr-qc/9709002.

\bibitem{Madore:2002fk}
J.~Madore, M.~Maceda, and D.C. Robinson.
\newblock {Fuzzy pp waves}.
\newblock pages 135--176, 2002, hep-th/0207225.

\bibitem{Freidel:2001kb}
Laurent Freidel and Kirill Krasnov.
\newblock {The Fuzzy sphere star product and spin networks}.
\newblock {\em J.Math.Phys.}, 43:1737--1754, 2002, hep-th/0103070.

\bibitem{Majid:1988we}
S.~Majid.
\newblock {Hopf algebras for physicts at the Planck scale}.
\newblock {\em Class.Quant.Grav.}, 5:1587--1606, 1988.

\bibitem{Maggiore:1993zu}
Michele Maggiore.
\newblock {Quantum groups, gravity and the generalized uncertainty principle}.
\newblock {\em Phys.Rev.}, D49:5182--5187, 1994, hep-th/9305163.

\bibitem{Majid:2000fk}
Shahn Majid.
\newblock Quantum groups and noncommutative geometry.
\newblock {\em J. MATH. PHYS}, pages 3892--3942.

\bibitem{Major:1995yz}
Seth Major and Lee Smolin.
\newblock {Quantum deformation of quantum gravity}.
\newblock {\em Nucl. Phys.}, B473:267--290, 1996, gr-qc/9512020.

\bibitem{Smolin:2002sz}
Lee Smolin.
\newblock {Quantum gravity with a positive cosmological constant}.
\newblock 2002, hep-th/0209079.

\bibitem{Freidel:1998pt}
Laurent Freidel and Kirill Krasnov.
\newblock {Spin foam models and the classical action principle}.
\newblock {\em Adv.Theor.Math.Phys.}, 2:1183--1247, 1999, hep-th/9807092.

\bibitem{Penrose2}
Roger Penrose.
\newblock Angular momentum: an approach to combinatorial spacetime.
\newblock In T.~Bastin, editor, {\em Quantum Theory and Beyond}, pages
  151--180, Cambridge, U.K., 1971. Cambridge University Press.

\bibitem{E.Buffenoir:kx}
Eric Buffenoir and Philippe Roche.
\newblock Harmonic analysis on the quantum {L}orentz group.
\newblock q-alg/9710022v2.

\bibitem{Noui:2002ag}
Karim Noui and Philippe Roche.
\newblock {Cosmological deformation of Lorentzian spin foam models}.
\newblock {\em Class. Quant. Grav.}, 20:3175--3214, 2003, gr-qc/0211109.

\bibitem{Han:2010pz}
Muxin Han.
\newblock {4-dimensional Spin-foam Model with Quantum Lorentz Group}.
\newblock {\em J. Math. Phys.}, 52:072501, 2011, 1012.4216.

\bibitem{Fairbairn:2010cp}
Winston~J. Fairbairn and Catherine Meusburger.
\newblock {Quantum deformation of two four-dimensional spin foam models}.
\newblock {\em J.Math.Phys.}, 53:022501, 2012, 1012.4784.

\bibitem{Han:2011nx}
Muxin Han.
\newblock Cosmological constant in lqg vertex amplitude.
\newblock {\em Phys. Rev. D}, 84:064010, 2011, 1105.2212.

\bibitem{Snyder:1946qz}
Hartland~S. Snyder.
\newblock {Quantized space-time}.
\newblock {\em Phys.Rev.}, 71:38--41, 1947.

\bibitem{Chamseddine:1992yx}
Ali~H. Chamseddine, Giovanni Felder, and J.~Frohlich.
\newblock {Gravity in noncommutative geometry}.
\newblock {\em Commun.Math.Phys.}, 155:205--218, 1993, hep-th/9209044.

\bibitem{Jevicki:2000it}
Antal Jevicki, Mihail Mihailescu, and Sanjaye Ramgoolam.
\newblock {Noncommutative spheres and the AdS / CFT correspondence}.
\newblock {\em JHEP}, 0010:008, 2000, hep-th/0006239.

\bibitem{Moffat:2000gr}
J.W. Moffat.
\newblock {Noncommutative quantum gravity}.
\newblock {\em Phys.Lett.}, B491:345--352, 2000, hep-th/0007181.

\bibitem{Vacaru:2000yk}
Sergiu~I. Vacaru.
\newblock {Gauge and Einstein gravity from nonAbelian gauge models on
  noncommutative spaces}.
\newblock {\em Phys.Lett.}, B498:74--86, 2001, hep-th/0009163.

\bibitem{Cacciatori:2002ib}
S.~Cacciatori, A.H. Chamseddine, D.~Klemm, L.~Martucci, W.A. Sabra, et~al.
\newblock {Noncommutative gravity in two dimensions}.
\newblock {\em Class.Quant.Grav.}, 19:4029--4042, 2002, hep-th/0203038.

\bibitem{Cardella:2002pb}
Matteo~A. Cardella and Daniela Zanon.
\newblock {Noncommutative deformation of four-dimensional Einstein gravity}.
\newblock {\em Class.Quant.Grav.}, 20:L95--L104, 2003, hep-th/0212071.

\bibitem{Vassilevich:2004ym}
Dimitri~V. Vassilevich.
\newblock {Quantum noncommutative gravity in two dimensions}.
\newblock {\em Nucl.Phys.}, B715:695--712, 2005, hep-th/0406163.

\bibitem{Buric:2006di}
Maja Buric, Theodoros Grammatikopoulos, John Madore, and George Zoupanos.
\newblock {Gravity and the structure of noncommutative algebras}.
\newblock {\em JHEP}, 0604:054, 2006, hep-th/0603044.

\bibitem{Abe:2002in}
Yasuhiro Abe and V.P. Nair.
\newblock {Noncommutative gravity: Fuzzy sphere and others}.
\newblock {\em Phys.Rev.}, D68:025002, 2003, hep-th/0212270.

\bibitem{Valtancoli:2003ve}
Paolo Valtancoli.
\newblock {Gravity on a fuzzy sphere}.
\newblock {\em Int.J.Mod.Phys.}, A19:361--370, 2004, hep-th/0306065.

\bibitem{Kurkcuoglu:2006iw}
Seckin Kurkcuoglu and Christian Saemann.
\newblock {Drinfeld Twist and General Relativity with Fuzzy Spaces}.
\newblock {\em Class.Quant.Grav.}, 24:291--312, 2007, hep-th/0606197.

\bibitem{Krajewski:1999bg}
Thomas Krajewski and Igor Pris.
\newblock {Hopf algebras from the quantum geometry point of view}.
\newblock 1999.

\bibitem{Grosse:2004yu}
Harald Grosse and Raimar Wulkenhaar.
\newblock {Renormalization of phi**4 theory on noncommutative R**4 in the
  matrix base}.
\newblock {\em Commun.Math.Phys.}, 256:305--374, 2005, hep-th/0401128.

\bibitem{Freidel:2005me}
Laurent Freidel and Etera Livine.
\newblock Effective 3d quantum gravity and non-commutative quantum field
  theory.
\newblock {\em Phys. Rev. Lett.}, 96:221301, 2006.

\bibitem{Szabo:2006wx}
Richard~J. Szabo.
\newblock {Symmetry, gravity and noncommutativity}.
\newblock {\em Class.Quant.Grav.}, 23:R199--R242, 2006, hep-th/0606233.

\bibitem{Livine:2008hz}
Etera~R. Livine.
\newblock The non-commutative geometry of matrix models: the spinfoam way.
\newblock 2008, 0811.1462.

\bibitem{albook}
Alain Connes.
\newblock {\em Noncommutative geometry}.
\newblock Academic Press, San Diego, CA, 1994.

\bibitem{AmelinoCamelia:2011bm}
Giovanni Amelino-Camelia, Laurent Freidel, Jerzy Kowalski-Glikman, and Lee
  Smolin.
\newblock {The principle of relative locality}.
\newblock 2011, 1101.0931.
\newblock * Temporary entry *.

\bibitem{Girelli:2003az}
Florian Girelli and Etera~R. Livine.
\newblock {Quantizing speeds with the cosmological constant}.
\newblock {\em Phys.Rev.}, D69:104024, 2004, gr-qc/0311032.

\bibitem{Bianchi:2010zs}
Eugenio Bianchi, Carlo Rovelli, and Francesca Vidotto.
\newblock {Towards Spinfoam Cosmology}.
\newblock {\em Phys. Rev.}, D82:084035, 2010, 1003.3483.

\bibitem{Bianchi:2011ym}
Eugenio Bianchi, Thomas Krajewski, Carlo Rovelli, and Francesca Vidotto.
\newblock {Cosmological constant in spinfoam cosmology}.
\newblock {\em Phys. Rev.}, D83:104015, 2011, 1101.4049.

\bibitem{Vidotto:2011qa}
Francesca Vidotto.
\newblock {Many-nodes/many-links spinfoam: the homogeneous and isotropic case}.
\newblock {\em Class. Quant Grav.}, 28(245005), 2011, 1107.2633.

\bibitem{Rennert:2013pf}
Julian Rennert and David Sloan.
\newblock {Towards Anisotropic Spinfoam Cosmology}.
\newblock 2013, 1304.6688.

\bibitem{Alesci:2013xd}
Emanuele Alesci and Francesco Cianfrani.
\newblock {Quantum-Reduced Loop Gravity: Cosmology}.
\newblock 2013, 1301.2245.

\bibitem{Eliade1957}
Mircea Eliade.
\newblock {\em The sacred and the profane: The nature of religion}.
\newblock Houghton Mifflin Harcourt, 1957.

\bibitem{Sand1956}
Archimedes.
\newblock The sand reckoner.
\newblock In James~R. Newman, editor, {\em The World of Mathematics}. Simon \&
  Schuste, 1956.

\bibitem{Sirach}
Joshua ben Sira.
\newblock Ecclesiasticus.
\newblock In {\em The Bible}. 2nd century BC.

\bibitem{Infinity2011}
Carlo Rovelli.
\newblock Some considerations on infinity in physics.
\newblock In M.~Heller and W.H. Woodin, editors, {\em Infinity. New Research
  Frontiers}. Cambridge University Press, 2011.

\end{thebibliography}
\end{document}